\documentclass[english]{revtex4-1}
\usepackage[T1]{fontenc}
\usepackage[latin9]{inputenc}
\usepackage{babel}
\usepackage{textcomp}
\usepackage{amsmath}
\usepackage{graphicx}
\usepackage[unicode=true,pdfusetitle,
 bookmarks=true,bookmarksnumbered=false,bookmarksopen=false,
 breaklinks=false,pdfborder={0 0 1},backref=false,colorlinks=false]
 {hyperref}
\begin{document}
\title{Effective potential and dynamical symmetry breaking up to five loops
in a massless abelian Higgs model}
\author{A. G. Quinto}
\email{andresarturogomezquinto@mail.uniatlantico.edu.co}

\author{R. Vega Monroy}
\email{ricardovega@mail.uniatlantico.edu.co}

\affiliation{Facultad de Ciencias Básicas, Universidad del Atlántico Km. 7, Via
a Pto. Colombia, Barranquilla, Colombia}
\author{A. F. Ferrari}
\email{alysson.ferrari@ufabc.edu.br}

\affiliation{Centro de Ciências Naturais e Humanas, Universidade Federal do ABC--
UFABC, Rua Santa Adélia, 166, 09210-170, Santo André, SP, Brazil}
\author{A. C. Lehum}
\email{lehum@ufpa.br}

\affiliation{Faculdade de Física, Universidade Federal do Pará, 66075-110, Belém,
Pará, Brazil}
\begin{abstract}
In this paper, we investigate the application of the Renormalization
Group Equation (RGE) in the determination of the effective potential
and the study of Dynamical Symmetry Breaking (DSB) in a massless Abelian
Higgs (AH) model with an $N$-component complex scalar field in $(3+1)$
dimensional spacetime. The classical Lagrangian of this model has
scale invariance, which can be broken by radiative corrections to
the effective potential. It is possible to calculate the effective
potential using the RGE and the renormalization group functions that
are obtained directly from loop calculations of the model and, using
the leading logs approximation, information about higher loop orders
can be included in the effective potential thus obtained. To show
this, we use the renormalization group functions reported in the literature,
obtained with a four loop calculation, and obtain a five loop approximation
to the effective potential, in doing so, we have to properly take
into account the fact that the model has multiple scales, and convert
the functions that were originally calculated in the minimal subtraction
(MS) renormalization scheme to another scheme which is adequate for
the RGE method. This result is then used to study the DSB, and we
present evidence for a rich structure of classical vacua, depending
on the value of gauge coupling constant and number of scalar fields,
which are considered as free parameters.
\end{abstract}
\maketitle

\section{Introduction}

The Abelian Higgs (AH) model is one of the most fundamental field
theories in both condensed matter and particle physics. As an example,
it is the prime textbook example for the superconducting transition
and the Anderson-Higgs mechanism\,\citep{ZinnJustin1996,Peskin2018,Altland2010a,Herbut2007}.
The AH model features a complex scalar field coupled to an $U(1)$
gauge field, and it displays two distinct phases separated by a sharp
transition: the symmetric phase and the phase with spontaneously broken
symmetry. In the context of superconductors, the symmetric phase is
related to the normal metallic state and the broken one to the superconducting
Meissner state. The transition between spontaneously broken and symmetric
phases is characterized by a dimensionful parameter that serves as
an order parameter.

Starting from a classical scale invariant Lagrangian, Coleman and
Weinberg (CW) demonstrated in\,\citep{Coleman1973} how the order
parameter could be generated by radiative corrections, i.e., the spontaneous
symmetry breaking can occur as a dynamical mechanism, the radiative
corrections being entirely responsible for the appearance of the nontrivial
minima of the effective potential. This Dynamical Symmetry Breaking
(DSB) is a key concept that has many applications in particle physics\,\citep{Elias2003,Chishtie2006,Chishtie2011,Steele2013}
and condensate matter systems\,\citep{Liu2013,Uchino2014,Burmistrov2020,Chodos1994,A.G.Quinto2021}.

In order to study the CW mechanism, we need to calculate the effective
potential, a powerful tool to explore many aspects of the low-energy
sector of a quantum field theory. In many cases, the one-loop approximation
is good enough, but it can be improved by adding higher order contributions
to the loop expansion. A standard tool for improving a perturbative
calculation is the Renormalization Group Equation (RGE), which, together
with a reorganization of the perturbative series in terms of leading
logs, have been shown to be very effective in several instances\,\citep{AHMADY2003221,PhysRevD.72.037902,Souza2020,Quinto2016,Dias2014,CHISHTIE2007,A.G.Quinto2021}.
We refer the reader to section 3 in\,\citep{Quinto2016} for a short
review of the method, and\,\citep{Elias2003,Chishtie2006,Chishtie2011,Steele2013}
for some of the interesting results that have been reported with the
use of the RG improvement, in the context of a scale-invariant approximation
of the Standard Model.

In this work we study the behavior of effective potential in a massless
AH model with $N$-component complex scalar field in $(3+1)$ dimensional
space-time, which is scale invariant at the classical level. We observed
that the effective potential, computed up to five loops, leads to
an interesting phase structure arising from DSB. This result was achieved
by the use of RGE with the help of renormalization group functions,
$\widetilde{\beta}$ and $\widetilde{\gamma}$, which were calculated
up to four loops in the minimal subtraction (MS) scheme in\,\citep{Ihrig2019}.
From these renormalization group functions, we need to obtain the
corresponding functions in a different renormalization scheme, which
we call CW scheme, using the multi-scales techniques reported in\,\citep{Chishtie2008}.

This paper is organized as follows: in Sec.\,\ref{sec:beta funtion for abelian higgs model},
we present our model, together with the renormalization group functions
found in the literature. In Section\,\ref{sec:-Beta-function-in-the-CW-scheme}
we obtain the corresponding functions in the CW scheme and we use
them in Section\,\ref{sec:effective potential up to four loops}
for the calculation of the effective potential using the RGE approach.
This effective potential is used in Section\,\ref{sec:Dynamical-symmetric-breaking}
to study different aspects of the DSB in our model. Section\,\ref{sec:Conclusion}
presents our conclusions and perspectives.

\section{\label{sec:beta funtion for abelian higgs model}the massless abelian
higgs model in the MS scheme and its corresponding $\widetilde{\beta}$
and $\widetilde{\gamma}$ functions}

We start with the $N$-component massless Abelian Higgs (AH) model
defined in $d$-dimensional Euclidean space-time by the Lagrangian

\begin{align}
\mathcal{L} & =\left|D_{\mu}\phi\right|^{2}+\frac{1}{4}F_{\mu v}^{2}+\lambda\left(|\phi|^{2}\right)^{2}+\mathcal{L}_{\mathrm{gf}},\label{eq:massless abelian Higgs model}
\end{align}
where $\phi=\left(\phi_{1},\ldots,\phi_{N}\right)$ describes the
$N$-component complex scalar field with quartic self-interaction
$\lambda$. This scalar is minimally coupled to an $U(1)$ gauge field
$A_{\mu}$ via covariant derivative $D_{\mu}=\partial_{\mu}-ieA_{\mu}$,
$e$ being the analogous to the ``electric charge''. The field strength
tensor is defined as $F_{\mu v}=\partial_{\mu}A_{v}-\partial_{\nu}A_{\mu}$
and the gauge-fixing Lagrangian is $\mathcal{L}_{\mathrm{gf}}=-\frac{1}{2\xi}\left(\partial_{\mu}A^{\mu}\right)^{2}$,
$\xi$ being the gauge-fixing parameter. For the case of a single
complex scalar field, $N=1$, and in three spatial dimensions, this
model is used to describe transitions on superconductors\,\citep{Ginzburg1950}
and liquid crystals\,\citep{Halperin1974}.

The renormalized Lagrangian of the massless AH model is

\begin{align}
\mathcal{L}^{\prime} & =Z_{\phi}\left|D_{\mu}\phi\right|^{2}+Z_{\phi^{4}}\lambda\widetilde{\mu}^{\epsilon}\left(|\phi|^{2}\right)^{2}+\frac{Z_{A}}{4}F_{\mu v}^{2}-\frac{1}{2\xi}\left(\partial_{\mu}A^{\mu}\right)^{2},\label{eq:renormalized lagrangiam}
\end{align}
where $D_{\mu}\phi=\left(\partial_{\mu}-ie\widetilde{\mu}^{\epsilon/2}A_{\mu}\right)\phi$
and $\widetilde{\mu}$ is a mass scale introduced by the (dimensional)
regularization scheme\,\citep{Collins1984,Peskin2018,Altland2010}.
The wave-function renormalization constants $Z_{\phi}$ and $Z_{A}$
relate the bare and the renormalized fields in the Lagrangian through
$\phi_{0}=Z_{\phi}^{1/2}\phi$ and $A_{0,\mu}=Z_{A}^{1/2}A_{\mu}$.
Also, we obtain the relations between bare and renormalized coupling
constants as
\begin{align}
\alpha & \equiv e^{2}=e_{0}^{2}\widetilde{\mu}^{-\epsilon}Z_{A},\\
\lambda & =\lambda_{0}\widetilde{\mu}^{-\epsilon}Z_{\phi}^{2}Z_{\phi^{4}}^{-1},\\
\xi & =\xi_{0}Z_{A}^{-1}.
\end{align}
It is interesting to note that the gauge-fixing parameter is also
renormalized\,\citep{Collins1984} in this case, meaning it will
have a corresponding $\beta$ function.

In the MS scheme, the beta functions of the model are defined by 
\begin{align}
\widetilde{\beta}_{\alpha}=-\widetilde{\mu}\frac{d\alpha}{d\widetilde{\mu}},\; & \widetilde{\beta}_{\lambda}=-\widetilde{\mu}\frac{d\lambda}{d\widetilde{\mu}},\;\widetilde{\beta}_{\xi}=-\widetilde{\mu}\frac{d\xi}{d\widetilde{\mu}}.\label{eq:definition of beta functions}
\end{align}
These functions were computed in the Ref.\,\citep{Ihrig2019} up
to four loops, from which we quote the expressions below. The contributions
to $\widetilde{\beta}_{\alpha},\widetilde{\beta}_{\lambda}$ and $\widetilde{\beta}_{\xi}$,
can be cast as, for the gauge coupling constant,
\begin{align}
\widetilde{\beta}_{\alpha} & =\widetilde{\beta}_{\alpha}^{\left(2\right)}+\widetilde{\beta}_{\alpha}^{\left(3\right)}+\widetilde{\beta}_{\alpha}^{\left(4\right)}+\widetilde{\beta}_{\alpha}^{\left(5\right)},\label{eq:beta alpha}
\end{align}
where\begin{subequations} 
\begin{align}
\widetilde{\beta}_{\alpha}^{\left(2\right)} & =-\frac{N}{3}\alpha^{2},\\
\widetilde{\beta}_{\alpha}^{\left(3\right)} & =-2N\alpha^{3},\\
\widetilde{\beta}_{\alpha}^{\left(4\right)} & =\left(\frac{49N^{2}}{72}-\frac{29N}{8}\right)\alpha^{4}-\frac{1}{2}\left(N^{2}+N\right)\alpha^{3}\lambda+\frac{1}{8}\left(N^{2}+N\right)\alpha^{2}\lambda^{2},\\
\widetilde{\beta}_{\alpha}^{\left(5\right)} & =\left(\frac{3N}{16}-\frac{323N^{3}}{3888}+\left(\frac{451}{54}-\frac{38\zeta_{3}}{9}\right)N^{2}\right)\alpha^{5}+\left(\frac{5N^{3}}{72}-\frac{41N^{2}}{9}-\frac{37N}{8}\right)\alpha^{4}\lambda\nonumber \\
 & +\left(\frac{N^{3}}{24}+\frac{139N^{2}}{48}+\frac{137N}{48}\right)\alpha^{3}\lambda^{2}-\left(\frac{5N^{3}}{48}+\frac{7N^{2}}{16}+\frac{N}{3}\right)\alpha^{2}\lambda^{3},
\end{align}
\end{subequations} for the scalar self-interaction,
\begin{align}
\widetilde{\beta}_{\lambda} & =\widetilde{\beta}_{\lambda}^{\left(2\right)}+\widetilde{\beta}_{\lambda}^{\left(3\right)}+\widetilde{\beta}_{\lambda}^{\left(4\right)}+\widetilde{\beta}_{\lambda}^{\left(5\right)},\label{eq:beta lambda}
\end{align}
where\begin{subequations} 
\begin{align}
\widetilde{\beta}_{\lambda}^{\left(2\right)} & =-6\alpha^{2}+6\alpha\lambda-\left(4+N\right)\lambda^{2},\\
\widetilde{\beta}_{\lambda}^{\left(3\right)} & =\left(\frac{14N}{3}+30\right)\alpha^{3}-\left(\frac{71N}{6}+\frac{29}{2}\right)\alpha^{2}\lambda-\left(4N+10\right)\alpha\lambda^{2}+\left(\frac{9N}{2}+\frac{21}{2}\right)\lambda^{3},\\
\widetilde{\beta}_{\lambda}^{\left(4\right)} & =\left(-45\zeta_{3}-\frac{7N^{2}}{18}+\left(\frac{203}{8}-27\zeta_{3}\right)N+\frac{367}{8}\right)\alpha^{4}+\left(\left(18\zeta_{3}-\frac{989}{8}\right)N-54\zeta_{3}-\frac{5N^{2}}{216}-\frac{889}{4}\right)\alpha^{3}\lambda,\nonumber \\
 & +\left(126\zeta_{3}+\frac{43N^{2}}{16}+\left(18\zeta_{3}+\frac{1749}{16}\right)N+\frac{1093}{8}\right)\alpha^{2}\lambda^{2}+\left(6\zeta_{3}+\left(\frac{25}{2}-6\zeta_{3}\right)N+\frac{29}{2}\right)\alpha\lambda^{3}\nonumber \\
 & +\left(-33\zeta_{3}-\frac{33N^{2}}{16}-\left(15\zeta_{3}+\frac{461}{16}\right)N-\frac{185}{4}\right)\lambda^{4},\\
\widetilde{\beta}_{\lambda}^{\left(5\right)} & =\left(504\zeta_{3}-390\zeta_{5}+\left(\frac{1}{81}-\frac{2\zeta_{3}}{9}\right)N^{3}+\left(28\zeta_{3}-\frac{\pi^{4}}{10}-\frac{55709}{1296}\right)N^{2}+\left(310\zeta_{5}-\frac{578\zeta_{3}}{3}-\frac{19\pi^{4}}{60}-\frac{13987}{36}\right)N\right.\nonumber \\
 & \left.-\frac{33\pi^{4}}{20}-\frac{12751}{16}\right)\alpha^{5}+\left(2095\zeta_{5}-\frac{1191\zeta_{3}}{2}+\left(\frac{19\zeta_{3}}{18}+\frac{67}{2592}\right)N^{3}+\left(20\zeta_{5}-\frac{77\zeta_{3}}{2}+\frac{\pi^{4}}{5}+\frac{12779}{162}\right)N^{2}\right.\nonumber \\
 & +\left.\left(\frac{431\zeta_{3}}{2}+305\zeta_{5}+\frac{13\pi^{4}}{10}+\frac{209}{12}\right)N+\frac{26\pi^{4}}{5}-\frac{19127}{96}\right)\alpha^{4}\lambda-\left(725\zeta_{3}-1520\zeta_{5}+\left(\frac{7\zeta_{3}}{6}+\frac{139}{1944}\right)N^{3}\right.\nonumber \\
 & \left.-\left(40\zeta_{5}-\frac{157\zeta_{3}}{3}-\frac{\pi^{4}}{60}+\frac{289817}{7776}\right)N^{2}-\left(1080\zeta_{5}-\frac{1813\zeta_{3}}{2}-\frac{5\pi^{4}}{12}+\frac{88871}{96}\right)N+\frac{6\pi^{4}}{5}-\frac{66851}{48}\right)\alpha^{3}\lambda^{2}\nonumber \\
 & +\left(\left(-\frac{27\zeta_{3}}{2}+40\zeta_{5}-\frac{\pi^{4}}{180}-\frac{6145}{48}\right)N^{2}-768\zeta_{3}-960\zeta_{5}-\left(\frac{3157\zeta_{3}}{6}+80\zeta_{5}+\frac{143\pi^{4}}{180}+\frac{25123}{24}\right)N\right.\nonumber \\
 & \left.-\frac{12\pi^{4}}{5}-\frac{18503}{16}\right)\alpha^{2}\lambda^{3}+\left(150\zeta_{5}-435\zeta_{3}+\left(\frac{29\zeta_{3}}{2}-\frac{\pi^{4}}{60}-\frac{377}{96}\right)N^{2}+\frac{103\pi^{4}}{60}-\frac{185}{12}\right.\nonumber \\
 & \left.+\left(50\zeta_{5}-\frac{269\zeta_{3}}{2}+\frac{7\pi^{4}}{10}-\frac{1403}{32}\right)N\right)\alpha\lambda^{4}+\left(\frac{583\zeta_{3}}{2}+465\zeta_{5}-\frac{5N^{3}}{96}+\left(\frac{63\zeta_{3}}{2}+20\zeta_{5}-\frac{\pi^{4}}{12}+\frac{395}{12}\right)N^{2}\right.\nonumber \\
 & \left.+\left(191\zeta_{3}+275\zeta_{5}-\frac{31\pi^{4}}{60}+\frac{10057}{48}\right)N-\frac{11\pi^{4}}{15}+\frac{24581}{96}\right)\lambda^{5},
\end{align}
\end{subequations} and finally, for the gauge parameter,
\begin{align}
\widetilde{\beta}_{\xi} & =\widetilde{\beta}_{\xi}^{\left(2\right)}+\widetilde{\beta}_{\xi}^{\left(3\right)}+\widetilde{\beta}_{\xi}^{\left(4\right)}+\widetilde{\beta}_{\xi}^{\left(5\right)},\label{eq:beta xi}
\end{align}
where\begin{subequations} 
\begin{align}
\widetilde{\beta}_{\xi}^{\left(2\right)} & =\frac{8N}{3}\alpha\xi,\\
\widetilde{\beta}_{\xi}^{\left(3\right)} & =16N\alpha^{2}\xi,\\
\widetilde{\beta}_{\xi}^{\left(4\right)} & =\left(29-\frac{49N}{9}\right)N\alpha^{3}\xi+\frac{17}{6}N\left(N+1\right)\alpha^{2}\lambda\xi-\frac{5}{12}N\left(N+1\right)\alpha\lambda^{2}\xi,\\
\widetilde{\beta}_{\xi}^{\left(5\right)} & =\left(\frac{1}{486}N\left(323N^{2}+72\left(228\zeta_{3}-451\right)N-729\right)\right)\alpha^{4}\xi-\frac{25}{288}N\left(5N^{2}-328N-333\right)\alpha^{3}\lambda\xi\nonumber \\
 & -\frac{3}{32}N\left(2N^{2}+139N+137\right)\alpha^{2}\lambda^{2}\xi+\frac{11}{192}N\left(5N^{2}+21N+16\right)\alpha\lambda^{3}\xi.
\end{align}
\end{subequations}

The anomalous dimensions are defined through the relation
\begin{align}
\widetilde{\gamma}_{\phi} & \equiv-\frac{\widetilde{\mu}}{\phi}\frac{d\phi}{d\widetilde{\mu}}=-\frac{\widetilde{\mu}}{Z_{\phi}}\frac{d}{d\widetilde{\mu}}Z_{\phi},\label{eq:definition of gamma phi}
\end{align}
and, up to four loops, the contributions to $\gamma_{\phi}$ read
as
\begin{align}
\widetilde{\gamma}_{\phi} & =\widetilde{\gamma}_{\phi}^{\left(1\right)}+\widetilde{\gamma}_{\phi}^{\left(2\right)}+\widetilde{\gamma}_{\phi}^{\left(3\right)}+\widetilde{\gamma}_{\phi}^{\left(4\right)},\label{eq:gamma phi}
\end{align}
where\begin{subequations} 
\begin{align}
\widetilde{\gamma}_{\phi}^{\left(1\right)} & =-2\alpha,\\
\widetilde{\gamma}_{\phi}^{\left(2\right)} & =-\alpha\xi+\frac{1}{12}\left(11N+9\right)\alpha^{2}+\frac{1}{4}\left(N+1\right)\lambda^{2},\\
\widetilde{\gamma}_{\phi}^{\left(3\right)} & =\frac{5}{4}\left(N+1\right)\alpha\lambda^{2}-\frac{1}{16}\left(N+1\right)\left(N+4\right)\lambda^{3}-\frac{1}{8}\left(24\text{\ensuremath{\zeta_{3}}}-13\right)\left(N+1\right)\alpha^{2}\lambda,\nonumber \\
 & +\left(\frac{1}{8}-3\text{\ensuremath{\zeta_{3}}}\left(N-1\right)+\frac{1}{432}N\left(5N+3267\right)\right)\alpha^{3}\\
\widetilde{\gamma}_{\phi}^{\left(4\right)} & =\left(\frac{125}{64}-\frac{5}{64}N^{3}+\frac{5}{8}N^{2}+\frac{85}{32}N\right)\lambda^{4}+\left(\frac{5}{6}-\zeta_{3}+\left(\frac{\zeta_{3}}{2}-\frac{19}{96}\right)N^{2}+\left(\frac{61}{96}-\frac{\zeta_{3}}{2}\right)N\right)\alpha\lambda^{3}\nonumber \\
 & +\left(\frac{63\zeta_{3}}{2}-45\zeta_{5}+\left(\frac{13}{5184}-\frac{\zeta_{3}}{36}\right)N^{3}+\left(\frac{9\zeta_{3}}{4}-\frac{\pi^{4}}{120}-\frac{1505}{432}\right)N^{2}+\left(\frac{231}{16}-\frac{87\zeta_{3}}{2}-\frac{\pi^{4}}{24}\right)N-\frac{\pi^{4}}{20}+\frac{133}{64}\right)\alpha^{4}\nonumber \\
 & +\left(\frac{345}{16}-4\zeta_{3}-20\zeta_{5}+\left(\frac{5\zeta_{3}}{3}-\frac{\pi^{4}}{180}+\frac{199}{144}\right)N^{2}+\left(-\frac{7\zeta_{3}}{3}-20\zeta_{5}+\frac{2\pi^{4}}{45}+\frac{413}{18}\right)N+\frac{\pi^{4}}{20}\right)\alpha^{3}\lambda\nonumber \\
 & +\left(\frac{11\zeta_{3}}{2}+5\zeta_{5}+\left(\frac{19\zeta_{3}}{12}-\frac{\pi^{4}}{120}-\frac{641}{288}\right)N^{2}+\left(\frac{85\zeta_{3}}{12}+5\zeta_{5}-\frac{\pi^{4}}{24}-\frac{179}{18}\right)N-\frac{\pi^{4}}{30}-\frac{247}{32}\right)\alpha^{2}\lambda^{2}.
\end{align}
\end{subequations}

The superscript present in the previous expressions denotes the aggregate
power of coupling constants. So, for instance, $\widetilde{\beta}_{\alpha}^{\left(2\right)}$
means the terms in $\widetilde{\beta}_{\alpha}$ which contain exactly
two powers of coupling constants.

The renormalization group functions presented in this section were
calculated in the MS scheme by\,\citep{Ihrig2019}. In order to use
the RGE improvement method, we need to convert these functions to
a different renormalization scheme\,\citep{Quinto2016,AHMADY2003221,Dias2010,PhysRevD.72.037902,Dias2014,CHISHTIE2007}.
This will be done in the next section.

\section{\label{sec:-Beta-function-in-the-CW-scheme}$\beta$ and $\gamma$
function in the CW scheme}

In this section we will use the renormalization group function obtained
in section\,\ref{sec:beta funtion for abelian higgs model} to calculate
the $\beta$ and $\gamma$ function in the CW scheme. We know the
effective potential will involve terms with logarithms, of the general
form
\begin{align}
L & =\ln\left(\frac{\sigma^{2}}{\mu^{2}}\right)\,\,\text{for CW scheme },\label{eq:CW scheme}\\
\widetilde{L} & =\ln\left(x\frac{\sigma^{2}}{\widetilde{\mu}^{2}}\right)\,\,\text{for MS scheme },\label{eq:MS echeme}
\end{align}
where $x$ is associated to some coupling constant present in the
model, and $\sigma$ is the classical value of one of the components
of $\phi$, the one which is shifted as $\phi_{i}\rightarrow\phi_{i}+\sigma$
in order to study the symmetry breaking (see details in Section\,\ref{sec:Dynamical-symmetric-breaking}).

We can obtain the relation between the renormalization group function
in the CW scheme from the knowledge of the corresponding function
in the MS scheme. This procedure is not straightforward because we
have multiple coupling constants and then we have to use the multi-scale
procedure described in\,\citep{Chishtie2008}. In order to do that,
we start with Eq.\,(\ref{eq:MS echeme}) applying to our model, i.e.,
$x=\lambda,\alpha,\xi$, then we get\begin{subequations} 
\begin{align}
\widetilde{L}_{1} & =\ln\left(\lambda\frac{\sigma^{2}}{k_{1}^{2}}\right),\\
\widetilde{L}_{2} & =\ln\left(\alpha\frac{\sigma^{2}}{k_{2}^{2}}\right),\\
\widetilde{L}_{3} & =\ln\left(\xi\frac{\sigma^{2}}{k_{3}^{2}}\right),
\end{align}
\end{subequations}where $k_{i}$ with $i=1,2,3$ are different scales.
Notice the gauge parameter appears here as if it were a coupling constant,
being dimensionless and having its own $\beta$ function as $\lambda$
and $\alpha$. If we compare $\widetilde{L}_{i}$ with Eq.\,(\ref{eq:CW scheme})
we obtain the following relations:
\begin{align}
k_{1} & =\lambda^{1/2}\mu,\nonumber \\
k_{2} & =\alpha^{1/2}\mu,\label{eq: multiscale relations}\\
k_{3} & =\xi^{1/2}\mu.\nonumber 
\end{align}

The renormalization group function in the CW scheme can be defined
as

\begin{align}
\beta_{\lambda}= & \mu\frac{d}{d\mu}\lambda=\mu\frac{\partial\lambda}{\partial k_{1}}\frac{dk_{1}}{d\mu}+\mu\frac{\partial\lambda}{\partial k_{2}}\frac{dk_{2}}{d\mu}+\mu\frac{\partial\lambda}{\partial k_{3}}\frac{dk_{3}}{d\mu},\label{eq: def beta lambda CW}\\
\beta_{\alpha}= & \mu\frac{d}{d\mu}\alpha=\mu\frac{\partial\alpha}{\partial k_{1}}\frac{dk_{1}}{d\mu}+\mu\frac{\partial\alpha}{\partial k_{2}}\frac{dk_{2}}{d\mu}+\mu\frac{\partial\alpha}{\partial k_{3}}\frac{dk_{3}}{d\mu},\label{eq:def beta alpha CW}\\
\beta_{\xi}= & \mu\frac{d}{d\mu}\xi=\mu\frac{\partial\xi}{\partial k_{1}}\frac{dk_{1}}{d\mu}+\mu\frac{\partial\xi}{\partial k_{2}}\frac{dk_{2}}{d\mu}+\mu\frac{\partial\xi}{\partial k_{3}}\frac{dk_{3}}{d\mu},\label{eq:def xi lambda CW}\\
\gamma_{\phi}= & \frac{\mu}{\phi}\frac{d}{d\mu}\phi=\frac{\mu}{\phi}\frac{\partial\phi}{\partial k_{1}}\frac{dk_{1}}{d\mu}+\frac{\mu}{\phi}\frac{\partial\phi}{\partial k_{2}}\frac{dk_{2}}{d\mu}+\frac{\mu}{\phi}\frac{\partial\phi}{\partial k_{3}}\frac{dk_{3}}{d\mu}.\label{eq:def gamma phi CW}
\end{align}
Now, if we use the relations Eq.\,(\ref{eq: multiscale relations})
in the last set of equations with the condition $k_{1}=k_{2}=k_{3}=\widetilde{\mu}$,
we get the final relation between the renormalization group functions
in the CW computed from MS scheme,\begin{subequations} 
\begin{align}
\beta_{\lambda} & =-\widetilde{\beta}_{\lambda}\left(3+\frac{\beta_{\text{\ensuremath{\lambda}}}}{2\lambda}+\frac{\beta_{\text{\ensuremath{\alpha}}}}{2\alpha}+\frac{\beta_{\text{\ensuremath{\xi}}}}{2\xi}\right),\\
\beta_{\alpha} & =-\widetilde{\beta}_{\alpha}\left(3+\frac{\beta_{\text{\ensuremath{\lambda}}}}{2\lambda}+\frac{\beta_{\text{\ensuremath{\alpha}}}}{2\alpha}+\frac{\beta_{\text{\ensuremath{\xi}}}}{2\xi}\right),\\
\beta_{\xi} & =-\widetilde{\beta}_{\xi}\left(3+\frac{\beta_{\text{\ensuremath{\lambda}}}}{2\lambda}+\frac{\beta_{\text{\ensuremath{\alpha}}}}{2\alpha}+\frac{\beta_{\text{\ensuremath{\xi}}}}{2\xi}\right),\\
\gamma_{\phi} & =-\widetilde{\gamma}_{\phi}\left(3+\frac{\beta_{\text{\ensuremath{\lambda}}}}{2\lambda}+\frac{\beta_{\text{\ensuremath{\alpha}}}}{2\alpha}+\frac{\beta_{\text{\ensuremath{\xi}}}}{2\xi}\right).
\end{align}
\end{subequations}Notice that the minus sign come from the definition
of the renormalization group function in the MS scheme (see Eqs.\,(\ref{eq:definition of beta functions})
and\,(\ref{eq:definition of gamma phi})).

We obtain the CW RG functions through an order by order comparison
of the previous expression. For example, for the lowest order, we
have\begin{subequations} 
\begin{align}
\gamma_{\phi}^{\left(1\right)} & =-3\widetilde{\gamma}_{\phi}^{\left(1\right)},\label{eq:gamma 1}\\
\beta_{\lambda}^{\left(2\right)} & =-3\widetilde{\beta}_{\lambda}^{\left(2\right)},\label{eq:beta lambda2}\\
\beta_{\alpha}^{\left(2\right)} & =-3\widetilde{\beta}_{\alpha}^{\left(2\right)},\label{eq:beta alpha 2}\\
\beta_{\xi}^{\left(2\right)} & =-3\widetilde{\beta}_{\xi}^{\left(2\right)},\label{eq:beta xi 2}
\end{align}
\end{subequations}where these relations are expected because at this
order the renormalization group function in the CW and MS are easily
related (see for example, the section 3 of the Ref.\,\citep{Quinto2016}
for more details).

For the next order, i.e, $\beta_{i}^{\left(3\right)}$ and $\gamma_{\phi}^{\left(2\right)}$,
the relation between the two schemes is more complex,\begin{subequations}
\begin{align}
\beta_{\lambda}^{\left(3\right)} & =-3\widetilde{\beta}_{\lambda}^{\left(3\right)}+3\widetilde{\beta}_{\lambda}^{\left(2\right)}\left(\frac{\widetilde{\beta}_{\lambda}^{\left(2\right)}}{2\lambda}+\frac{\widetilde{\beta}_{\alpha}^{\left(2\right)}}{2\alpha}+\frac{\widetilde{\beta}_{\xi}^{\left(2\right)}}{2\xi}\right),\\
\beta_{\alpha}^{\left(3\right)} & =-3\widetilde{\beta}_{\alpha}^{\left(3\right)}+3\widetilde{\beta}_{\alpha}^{\left(2\right)}\left(\frac{\widetilde{\beta}_{\lambda}^{\left(2\right)}}{2\lambda}+\frac{\widetilde{\beta}_{\alpha}^{\left(2\right)}}{2\alpha}+\frac{\widetilde{\beta}_{\xi}^{\left(2\right)}}{2\xi}\right),\\
\beta_{\xi}^{\left(3\right)} & =-3\widetilde{\beta}_{\xi}^{\left(3\right)}+3\widetilde{\beta}_{\xi}^{\left(2\right)}\left(\frac{\widetilde{\beta}_{\lambda}^{\left(2\right)}}{2\lambda}+\frac{\widetilde{\beta}_{\alpha}^{\left(2\right)}}{2\alpha}+\frac{\widetilde{\beta}_{\xi}^{\left(2\right)}}{2\xi}\right),\\
\gamma_{\phi}^{\left(2\right)} & =-3\widetilde{\gamma}_{\phi}^{\left(2\right)}+3\widetilde{\gamma}_{\phi}^{\left(1\right)}\left(\frac{\widetilde{\beta}_{\lambda}^{\left(2\right)}}{2\lambda}+\frac{\widetilde{\beta}_{\alpha}^{\left(2\right)}}{2\alpha}+\frac{\widetilde{\beta}_{\xi}^{\left(2\right)}}{2\xi}\right).
\end{align}
\end{subequations} For the order $\beta_{i}^{\left(4\right)}$ and
$\gamma_{\phi}^{\left(3\right)}$ we get\begin{subequations} 
\begin{align}
\beta_{\lambda}^{\left(4\right)} & =-3\widetilde{\beta}_{\lambda}^{\left(4\right)}+3\widetilde{\beta}_{\lambda}^{\left(3\right)}\left(\frac{\widetilde{\beta}_{\lambda}^{\left(2\right)}}{2\lambda}+\frac{\widetilde{\beta}_{\alpha}^{\left(2\right)}}{2\alpha}+\frac{\widetilde{\beta}_{\xi}^{\left(2\right)}}{2\xi}\right)-\widetilde{\beta}_{\lambda}^{\left(2\right)}\left(\frac{\beta_{\lambda}^{\left(3\right)}}{2\lambda}+\frac{\beta_{\alpha}^{\left(3\right)}}{2\alpha}+\frac{\beta_{\xi}^{\left(3\right)}}{2\xi}\right),\\
\beta_{\alpha}^{\left(4\right)} & =-3\widetilde{\beta}_{\alpha}^{\left(4\right)}+3\widetilde{\beta}_{\alpha}^{\left(3\right)}\left(\frac{\widetilde{\beta}_{\lambda}^{\left(2\right)}}{2\lambda}+\frac{\widetilde{\beta}_{\alpha}^{\left(2\right)}}{2\alpha}+\frac{\widetilde{\beta}_{\xi}^{\left(2\right)}}{2\xi}\right)-\widetilde{\beta}_{\alpha}^{\left(2\right)}\left(\frac{\beta_{\lambda}^{\left(3\right)}}{2\lambda}+\frac{\beta_{\alpha}^{\left(3\right)}}{2\alpha}+\frac{\beta_{\xi}^{\left(3\right)}}{2\xi}\right),\\
\beta_{\xi}^{\left(4\right)} & =-3\widetilde{\beta}_{\xi}^{\left(4\right)}+3\widetilde{\beta}_{\xi}^{\left(3\right)}\left(\frac{\widetilde{\beta}_{\lambda}^{\left(2\right)}}{2\lambda}+\frac{\widetilde{\beta}_{\alpha}^{\left(2\right)}}{2\alpha}+\frac{\widetilde{\beta}_{\xi}^{\left(2\right)}}{2\xi}\right)-\widetilde{\beta}_{\xi}^{\left(2\right)}\left(\frac{\beta_{\lambda}^{\left(3\right)}}{2\lambda}+\frac{\beta_{\alpha}^{\left(3\right)}}{2\alpha}+\frac{\beta_{\xi}^{\left(3\right)}}{2\xi}\right),\\
\gamma_{\phi}^{\left(3\right)} & =-3\widetilde{\gamma}_{\phi}^{\left(3\right)}+3\widetilde{\gamma}_{\phi}^{\left(2\right)}\left(\frac{\widetilde{\beta}_{\lambda}^{\left(2\right)}}{2\lambda}+\frac{\widetilde{\beta}_{\alpha}^{\left(2\right)}}{2\alpha}+\frac{\widetilde{\beta}_{\xi}^{\left(2\right)}}{2\xi}\right)-\widetilde{\gamma}_{\alpha}^{\left(1\right)}\left(\frac{\beta_{\lambda}^{\left(3\right)}}{2\lambda}+\frac{\beta_{\alpha}^{\left(3\right)}}{2\alpha}+\frac{\beta_{\xi}^{\left(3\right)}}{2\xi}\right).
\end{align}
\end{subequations} And finally, for the order $\beta_{i}^{\left(5\right)}$and
$\gamma_{\phi}^{\left(4\right)}$,\begin{subequations}

\begin{align}
\beta_{\lambda}^{\left(5\right)} & =-3\widetilde{\beta}_{\lambda}^{\left(5\right)}+3\widetilde{\beta}_{\lambda}^{\left(4\right)}\left(\frac{\widetilde{\beta}_{\lambda}^{\left(2\right)}}{2\lambda}+\frac{\widetilde{\beta}_{\alpha}^{\left(2\right)}}{2\alpha}+\frac{\widetilde{\beta}_{\xi}^{\left(2\right)}}{2\xi}\right)-\widetilde{\beta}_{\lambda}^{\left(3\right)}\left(\frac{\beta_{\lambda}^{\left(3\right)}}{2\lambda}+\frac{\beta_{\alpha}^{\left(3\right)}}{2\alpha}+\frac{\beta_{\xi}^{\left(3\right)}}{2\xi}\right)\nonumber \\
 & -\widetilde{\beta}_{\lambda}^{\left(2\right)}\left(\frac{\beta_{\lambda}^{\left(4\right)}}{2\lambda}+\frac{\beta_{\alpha}^{\left(4\right)}}{2\alpha}+\frac{\beta_{\xi}^{\left(4\right)}}{2\xi}\right),\\
\beta_{\alpha}^{\left(5\right)} & =-3\widetilde{\beta}_{\alpha}^{\left(5\right)}+3\widetilde{\beta}_{\alpha}^{\left(4\right)}\left(\frac{\widetilde{\beta}_{\lambda}^{\left(2\right)}}{2\lambda}+\frac{\widetilde{\beta}_{\alpha}^{\left(2\right)}}{2\alpha}+\frac{\widetilde{\beta}_{\xi}^{\left(2\right)}}{2\xi}\right)-\widetilde{\beta}_{\alpha}^{\left(3\right)}\left(\frac{\beta_{\lambda}^{\left(3\right)}}{2\lambda}+\frac{\beta_{\alpha}^{\left(3\right)}}{2\alpha}+\frac{\beta_{\xi}^{\left(3\right)}}{2\xi}\right)\nonumber \\
 & -\widetilde{\beta}_{\alpha}^{\left(2\right)}\left(\frac{\beta_{\lambda}^{\left(4\right)}}{2\lambda}+\frac{\beta_{\alpha}^{\left(4\right)}}{2\alpha}+\frac{\beta_{\xi}^{\left(4\right)}}{2\xi}\right),\\
\beta_{\xi}^{\left(5\right)} & =-3\widetilde{\beta}_{\xi}^{\left(5\right)}+3\widetilde{\beta}_{\xi}^{\left(4\right)}\left(\frac{\widetilde{\beta}_{\lambda}^{\left(2\right)}}{2\lambda}+\frac{\widetilde{\beta}_{\alpha}^{\left(2\right)}}{2\alpha}+\frac{\widetilde{\beta}_{\xi}^{\left(2\right)}}{2\xi}\right)-\widetilde{\beta}_{\xi}^{\left(3\right)}\left(\frac{\beta_{\lambda}^{\left(3\right)}}{2\lambda}+\frac{\beta_{\alpha}^{\left(3\right)}}{2\alpha}+\frac{\beta_{\xi}^{\left(3\right)}}{2\xi}\right)\nonumber \\
 & -\widetilde{\beta}_{\xi}^{\left(2\right)}\left(\frac{\beta_{\lambda}^{\left(4\right)}}{2\lambda}+\frac{\beta_{\alpha}^{\left(4\right)}}{2\alpha}+\frac{\beta_{\xi}^{\left(4\right)}}{2\xi}\right),\\
\gamma_{\phi}^{\left(4\right)} & =-3\widetilde{\gamma}_{\phi}^{\left(4\right)}+3\widetilde{\gamma}_{\phi}^{\left(3\right)}\left(\frac{\widetilde{\beta}_{\lambda}^{\left(2\right)}}{2\lambda}+\frac{\widetilde{\beta}_{\alpha}^{\left(2\right)}}{2\alpha}+\frac{\widetilde{\beta}_{\xi}^{\left(2\right)}}{2\xi}\right)-\widetilde{\gamma}_{\alpha}^{\left(2\right)}\left(\frac{\beta_{\lambda}^{\left(3\right)}}{2\lambda}+\frac{\beta_{\alpha}^{\left(3\right)}}{2\alpha}+\frac{\beta_{\xi}^{\left(3\right)}}{2\xi}\right)\nonumber \\
 & -\widetilde{\gamma}_{\alpha}^{\left(1\right)}\left(\frac{\beta_{\lambda}^{\left(4\right)}}{2\lambda}+\frac{\beta_{\alpha}^{\left(4\right)}}{2\alpha}+\frac{\beta_{\xi}^{\left(4\right)}}{2\xi}\right).
\end{align}

\end{subequations} Now, with the four-loop renormalization group
functions written in the CW scheme, in the next section we will compute
the effective potential and study the CW mechanism by using the RGE
in the leading log approximation up to five loops.

\section{\label{sec:effective potential up to four loops} effective potential
in the leading logs approximation}

The main object we shall be interested in studying is the effective
potential. In order to compute this object, we consider a shift in
the $N$-th component of $\phi$ in\,(\ref{eq:massless abelian Higgs model}),
\begin{align}
\phi_{N} & =\phi_{N}^{q}+\sigma,\label{eq:background field}
\end{align}
where $\phi_{N}^{q}=\frac{1}{\sqrt{2}}\left(\phi_{1N}+i\phi_{2N}\right)$
with $\phi_{1N}$ and $\phi_{2N}$ being two real scalar fields and
$\sigma$ is a constant expectation value of scalar field, called
\emph{background field}. This scalar field has the same properties
of $\phi_{N}$. If we substitute\,(\ref{eq:background field}) into\,(\ref{eq:massless abelian Higgs model})
we can find the effective potential in the classical approximation
\begin{align}
V_{\mathrm{eff}}^{\left(0\ell\right)}\left(\sigma\right) & =\frac{1}{4}\lambda\sigma^{4}.\label{eq:Classicla effective potential}
\end{align}

It is easy to see that $\sigma=0$ is the minimum of $V_{\mathrm{eff}}^{(0\ell)}$,
so there is no spontaneous symmetry breaking at classical level. Our
aim is to compute loop corrections to $V_{\mathrm{eff}}\left(\sigma\right)$
in order to understand if these corrections are capable to induce
a spontaneous symmetry breaking and the corresponding generation of
mass as given below

\begin{align}
m_{\phi_{1N}}^{2} & =\frac{3}{2}\lambda\sigma^{2},\\
m_{\phi_{2N}}^{2} & =\frac{1}{2}\lambda\sigma^{2}-\xi m_{A}^{2},\\
m_{A}^{2} & =\frac{1}{2}e^{2}\sigma^{2}.
\end{align}
Notice that the gauge dependence on the mass of $\phi_{2N}$ is a
consequence of a $R_{\xi}$-gauge, $\mathcal{L}_{\mathrm{gf}}=-\frac{1}{2\xi}\left(\partial_{\mu}A^{\mu}-\xi e\sigma\phi_{2N}\right)^{2}$.
Actually, in the spontaneously symmetry broken phase, the $\phi_{2N}$
degree of freedom is absorbed by the photon, which becomes massive.

As discussed in\,\citep{Quinto2016,A.G.Quinto2021}, the knowledge
of $V_{\mathrm{eff}}\left(\sigma\right)$ is sufficient for investigating
the dynamical breaking of gauge symmetry. We will be able to calculate
it by using an ansatz for the RGE, motivated by dimensional analysis,
together with the renormalization group functions for the model found
in section\,\ref{sec:-Beta-function-in-the-CW-scheme}. Specifically,
we shall use for $V_{\mathrm{eff}}\left(\sigma\right)$ the ansatz
\begin{align}
V_{\mathrm{eff}}\left(\sigma_{cl}\right) & =\sigma^{4}S_{\mathrm{eff}}\left(\sigma,\lambda,\alpha,\xi,L\right),\label{eq:ansatz for the effective potential}
\end{align}
where
\begin{align}
S_{\mathrm{eff}}\left(\sigma;\lambda,\alpha,\xi,L\right) & =A\left(\lambda,\alpha,\xi\right)+B\left(\lambda,\alpha,\xi\right)L+C\left(\lambda,\alpha,\xi\right)L^{2}+D\left(\lambda,\alpha,\xi\right)L^{3}+\cdots,\label{eq:effective action}
\end{align}
and $A,B,C,$ and $D$ are defined as power series in the coupling
constants $\lambda,\alpha,\xi$, and $L$ is defined in\,(\ref{eq:CW scheme}).
This ansatz follows from the conformal invariance at the tree-level,
leading to the fact that we can have only one type of logarithm appearing
in the quantum corrections. Comparison with\,(\ref{eq:Classicla effective potential})
show us that
\begin{align}
A\left(\lambda,\alpha,\xi\right) & =A^{\left(1\right)}=\frac{1}{4}\lambda.\label{eq:values for A}
\end{align}

Now, we need to calculate the $L$ dependent pieces of $V_{\mathrm{eff}}\left(\sigma\right)$,
involving $B,C,D$. For this, we need to use the RGE, in order to
obtain this equation we start with
\begin{align}
V_{\mathrm{eff}}^{0}\left(\sigma;\lambda_{0},\alpha_{0},\xi_{0}\right) & =V_{\mathrm{eff}}\left(\sigma;\lambda\left(\mu\right),\alpha\left(\mu\right),\xi\left(\mu\right),L\right),\label{eq:Vzero}
\end{align}
where $V_{\mathrm{eff}}^{0}$ is independent of mass scale $\mu$.
By deriving Eq.\,(\ref{eq:Vzero}) with respect to $\mu$,
\begin{align}
0 & =\left(\mu\frac{\partial}{\partial\mu}+\mu\frac{d\lambda}{d\mu}\frac{\partial}{\partial\lambda}+\mu\frac{d\alpha}{d\mu}\frac{\partial}{\partial\alpha}+\mu\frac{d\xi}{d\mu}\frac{\partial}{\partial\xi}+\mu\frac{d\sigma}{d\mu}\frac{\partial}{\partial\sigma}\right)V_{\mathrm{eff}}\left(\sigma;\lambda,\alpha,\xi,L\right),
\end{align}
and using Eqs.\,(\ref{eq: def beta lambda CW}) - (\ref{eq:def gamma phi CW}),
we have
\begin{align}
0 & =\left(\mu\frac{\partial}{\partial\mu}+\beta_{\lambda}\frac{\partial}{\partial\lambda}+\beta_{\alpha}\frac{\partial}{\partial\alpha}+\beta_{\xi}\frac{\partial}{\partial\xi}+\gamma_{\phi}\sigma\frac{\partial}{\partial\sigma}\right)V_{\mathrm{eff}}\left(\sigma;\lambda,\alpha,\xi,L\right),\label{eq:RGE in function of Veff}
\end{align}
finally, inserting Eq.\,(\ref{eq:ansatz for the effective potential})
into\,(\ref{eq:RGE in function of Veff}), we obtain an alternative
form for the RGE,

\begin{align}
\left[2\left(-1+\gamma_{\phi}\right)\partial_{L}+\beta_{\lambda}\partial_{\lambda}+\beta_{\alpha}\partial_{\alpha}+\beta_{\xi}\partial_{\xi}+4\gamma_{\phi}\right]S_{\mathrm{eff}}\left(\sigma;\lambda,\alpha,\xi,L\right) & =0,\label{eq:RGE}
\end{align}
were we used the notation $\partial_{x}\equiv\frac{\partial}{\partial x}$.

Inserting the ansatz\,(\ref{eq:effective action}) in\,(\ref{eq:RGE}),
and separating the resulting expression by orders of $L$, we obtain
a series of equations, 
\begin{align}
2\left(-1+\gamma_{\phi}\right)B\left(\lambda,\alpha,\xi\right)+\beta_{\lambda}\partial_{\lambda}A\left(\lambda,\alpha,\xi\right)+4\gamma_{\phi}A\left(\lambda,\alpha,\xi\right) & =0,\label{eq: Equation fo B}\\
2\left(-1+\gamma_{\phi}\right)C\left(\lambda,\alpha,\xi\right)+\left\{ \beta_{\lambda}\partial_{\lambda}+\beta_{\alpha}\partial_{\alpha}+\beta_{\xi}\partial_{\xi}+4\gamma_{\phi}\right\} B\left(\lambda,\alpha,\xi\right) & =0,\label{eq:Equation fo C}\\
2\left(-1+\gamma_{\phi}\right)D\left(\lambda,\alpha,\xi\right)+\left\{ \beta_{\lambda}\partial_{\lambda}+\beta_{\alpha}\partial_{\alpha}+\beta_{\xi}\partial_{\xi}+4\gamma_{\phi}\right\} C\left(\lambda,\alpha,\xi\right) & =0.\label{eq:Equation fo D}
\end{align}
As we can see in\,(\ref{eq: Equation fo B}) the function $A$ is
only dependent of the coupling $\lambda$, see Eq.\,(\ref{eq:values for A}),
for this reason the others beta functions were dropped.

We now consider that all functions appearing in Eq.\,(\ref{eq: Equation fo B})
are defined as series in powers of the couplings,

\begin{align}
-2\left(B^{\left(1\right)}+B^{\left(2\right)}+B^{\left(3\right)}+\ldots\right)+2\left(\gamma_{\phi}^{\left(1\right)}+\gamma_{\phi}^{\left(2\right)}+\ldots\right)\left(B^{\left(1\right)}+B^{\left(2\right)}+B^{\left(3\right)}+\ldots\right)\nonumber \\
+\left(\beta_{\lambda}^{\left(2\right)}+\beta_{\lambda}^{\left(3\right)}+\ldots\right)\partial_{\lambda}A^{\left(1\right)}+4\left(\gamma_{\phi}^{\left(1\right)}+\gamma_{\phi}^{\left(2\right)}+\ldots\right)A^{\left(1\right)} & =0,\label{eq: B equations order by order}
\end{align}
where the numbers in the superscripts denote the power of global coupling
constant of each term. Since all terms of the previous equation start
at order $\mathcal{O}\left(x^{2}\right)$, except the first, we conclude
that $B^{\left(1\right)}=0$, and obtain the relation
\begin{align}
B^{\left(2\right)} & =\frac{1}{8}\beta_{\lambda}^{\left(2\right)}+\frac{1}{2}\lambda\gamma_{\phi}^{\left(1\right)}=-\frac{3}{8}\widetilde{\beta}_{\lambda}^{\left(2\right)}-\frac{3}{2}\lambda\widetilde{\gamma}_{\phi}^{\left(1\right)}.
\end{align}
This last equation fixes the coefficients of $B^{\left(2\right)}$
in terms of (known) coefficients of $\widetilde{\beta}_{\lambda}^{\left(2\right)}$
and $\widetilde{\gamma}_{\phi}^{\left(1\right)}$, in the following
form,

\begin{align}
B^{\left(2\right)}= & \frac{3}{8}\left(6\alpha^{2}+2\alpha\lambda+(N+4)\lambda^{2}\right).
\end{align}
If we repeat the same procedure done in order to obtain $B^{\left(2\right)},$
we can find the others $B$\textasciiacute s with the helps of $\beta$\textasciiacute s
presents in\,(\ref{eq: B equations order by order}), as a results
we obtain
\begin{align}
B^{\left(3\right)} & =b_{1}\alpha\lambda\xi+b_{2}\alpha^{3}+b_{3}\alpha^{2}\lambda+b_{4}\alpha\lambda^{2}+b_{5}\lambda^{3}+b_{6}\alpha^{4}\lambda^{-1},\\
B^{\left(4\right)} & =b_{7}\alpha^{3}\xi+b_{8}\alpha^{4}+b_{9}\alpha\lambda^{2}\xi+b_{10}\alpha^{3}\lambda+b_{11}\alpha^{2}\lambda\xi+b_{12}\alpha^{2}\lambda^{2}+b_{13}\alpha\lambda^{3}+b_{14}\lambda^{4}\nonumber \\
 & +b_{15}\alpha^{5}\lambda^{-1}+b_{16}\alpha^{6}\lambda^{-2},\\
B^{\left(5\right)} & =b_{17}\lambda^{3}\alpha\xi+b_{18}\lambda^{3}\alpha^{2}+b_{19}\lambda^{2}\alpha^{3}+b_{20}\lambda^{2}\alpha^{2}\xi+b_{21}\lambda\alpha^{2}\xi^{2}+b_{22}\lambda\alpha^{4}+b_{23}\lambda\alpha^{3}\xi\nonumber \\
 & +b_{24}\alpha^{5}+b_{25}\alpha^{4}\xi+b_{26}\alpha\lambda^{4}+b_{27}\lambda^{5}+b_{28}\alpha^{5}\xi\lambda^{-1}+b_{29}\alpha^{6}\lambda^{-1}+b_{30}\alpha^{7}\lambda^{-2}+b_{31}\alpha^{8}\lambda^{-3}.
\end{align}
The coefficients $b_{1}$ to $b_{31}$ appearing in this equation
are defined in the appendix\,\ref{sec:All-coefficients of B}.

Now looking at Eq.\,(\ref{eq:Equation fo C}) expanded in powers
of the couplings,

\begin{align}
-4\left(C^{\left(1\right)}+C^{\left(2\right)}+C^{\left(3\right)}+\ldots\right)+4\left(\gamma_{\phi}^{\left(1\right)}+\gamma_{\phi}^{\left(2\right)}+\ldots\right)\left(C^{\left(1\right)}+C^{\left(2\right)}+C^{\left(3\right)}+\ldots\right)\nonumber \\
+\left[\left(\beta_{\lambda}^{\left(2\right)}+\beta_{\lambda}^{\left(3\right)}+\ldots\right)\partial_{\lambda}+\left(\beta_{\alpha}^{\left(2\right)}+\beta_{\alpha}^{\left(3\right)}+\ldots\right)\partial_{\alpha}+\left(\beta_{\xi}^{\left(2\right)}+\beta_{\xi}^{\left(3\right)}+\ldots\right)\partial_{\xi}\right]\sum_{n=2}^{4}B^{\left(n\right)}\nonumber \\
+4\left(\gamma_{\phi}^{\left(1\right)}+\gamma_{\phi}^{\left(2\right)}+\ldots\right)\sum_{n=2}^{4}B^{\left(n\right)} & =0,\label{eq:C order equation}
\end{align}
one may conclude that $C\left(\lambda,\alpha,\xi\right)$ starts at
order $C^{\left(3\right)}$, obtaining the relation,
\begin{align}
C^{\left(3\right)} & =\gamma_{\phi}^{\left(1\right)}B^{\left(2\right)}+\frac{1}{4}\left(\beta_{\lambda}^{\left(2\right)}\partial_{\lambda}+\beta_{\alpha}^{\left(2\right)}\partial_{\alpha}+\beta_{\xi}^{\left(2\right)}\partial_{\xi}\right)B^{\left(2\right)}\nonumber \\
 & =-3\widetilde{\gamma}_{\phi}^{\left(1\right)}B^{\left(2\right)}-\frac{3}{4}\left(\widetilde{\beta}_{\lambda}^{\left(2\right)}\partial_{\lambda}+\widetilde{\beta}_{\alpha}^{\left(2\right)}\partial_{\alpha}+\widetilde{\beta}_{\xi}^{\left(2\right)}\partial_{\xi}\right)B^{\left(2\right)},
\end{align}
from witch the coefficients of the form $x^{3}L^{2}$ of $S_{\mathrm{eff}}$
are calculated from known coefficients of the beta function, anomalous
dimension, and $B^{\left(2\right)}$. The end result is as follows,

\begin{align}
C^{\left(3\right)} & =\frac{9}{8}\left(N+15\right)\alpha^{3}+\frac{3}{16}\left(19N+78\right)\alpha^{2}\lambda-\frac{9}{16}\left(N+4\right)\alpha\lambda^{2}+\frac{9}{16}\left(N+4\right)^{2}\lambda^{3}.
\end{align}
If we repeat the same procedure, we can find the others $C$\textasciiacute s
with the helps of $\beta$\textasciiacute s and $\gamma'$s presents
in\,(\ref{eq:C order equation}), as a results we get,

\begin{align}
C^{\left(4\right)} & =c_{1}\alpha^{3}\xi+c_{2}\alpha^{4}+c_{3}\alpha^{3}\lambda+c_{4}\alpha^{2}\lambda^{2}+c_{5}\alpha^{2}\lambda\xi+c_{6}\alpha\lambda^{3}+c_{7}\alpha\lambda^{2}\xi+c_{8}\lambda^{4}\nonumber \\
 & +c_{9}\alpha^{5}\lambda^{-1}+c_{10}\alpha^{6}\lambda^{-2},\\
C^{\left(5\right)} & =c_{11}\lambda^{2}\alpha^{3}+c_{12}\lambda^{2}\alpha^{2}\xi+c_{13}\alpha^{5}+c_{14}\lambda^{3}\alpha^{2}+c_{15}\lambda^{3}\alpha\xi+c_{16}\alpha\lambda^{4}+c_{17}\alpha^{4}\xi\nonumber \\
 & +c_{18}\lambda\alpha^{2}\xi^{2}+c_{19}\lambda\alpha^{4}+c_{20}\lambda\alpha^{3}\xi+c_{21}\lambda^{5}+c_{22}\alpha^{5}\xi\lambda^{-1}+c_{23}\alpha^{6}\lambda^{-1}\nonumber \\
 & +c_{24}\alpha^{7}\lambda^{-2}+c_{25}\alpha^{8}\lambda^{-3}.
\end{align}
The coefficients $c_{1}$ to $c_{25}$ are presented in the appendix\,\ref{sec:All-coefficients-for C}.

Finally, looking at Eq.\,(\ref{eq:Equation fo D}) expanded in powers
of couplings,

\begin{align}
-6\left(D^{\left(1\right)}+D^{\left(2\right)}+D^{\left(3\right)}+D^{\left(4\right)}+\ldots\right)+6\left(\gamma_{\phi}^{\left(1\right)}+\gamma_{\phi}^{\left(2\right)}+\ldots\right)\left(D^{\left(1\right)}+D^{\left(2\right)}+D^{\left(3\right)}+D^{\left(4\right)}+\ldots\right)\nonumber \\
+\left[\left(\beta_{\lambda}^{\left(2\right)}+\beta_{\lambda}^{\left(3\right)}+\ldots\right)\partial_{\lambda}+\left(\beta_{\alpha}^{\left(2\right)}+\beta_{\alpha}^{\left(3\right)}+\ldots\right)\partial_{\alpha}+\left(\beta_{\xi}^{\left(2\right)}+\beta_{\xi}^{\left(3\right)}+\ldots\right)\partial_{\xi}\right]\left(C^{\left(3\right)}+C^{\left(4\right)}\right)\nonumber \\
+4\left(\gamma_{\phi}^{\left(1\right)}+\gamma_{\phi}^{\left(2\right)}+\ldots\right)\left(C^{\left(3\right)}+C^{\left(4\right)}\right) & =0,\label{eq:D order equation}
\end{align}
one may conclude that $D\left(\lambda,\alpha,\xi\right)$ starts at
order $D^{\left(4\right)}$, leading to the relation,
\begin{align}
D^{\left(4\right)} & =-\frac{4}{2}\widetilde{\gamma}_{\phi}^{\left(1\right)}C^{\left(3\right)}-\frac{1}{2}\left(\widetilde{\beta}_{\lambda}^{\left(2\right)}\partial_{\lambda}+\widetilde{\beta}_{\alpha}^{\left(2\right)}\partial_{\alpha}+\widetilde{\beta}_{\xi}^{\left(2\right)}\partial_{\xi}\right)C^{\left(3\right)},
\end{align}
from witch the coefficients of the form $x^{4}L^{3}$ of $S_{\mathrm{eff}}$
are calculated from the beta function, anomalous dimension, and $C^{\left(3\right)}$.
The end result is as follows,

\begin{align}
D^{\left(4\right)} & =\frac{9}{16}\left(N\left(N+42\right)+198\right)\alpha^{4}+\frac{1}{16}\left(N\left(19N+81\right)+18\right)\alpha^{3}\lambda+\frac{27}{16}\left(N+4\right)\left(4N+17\right)\alpha^{2}\lambda^{2}\nonumber \\
 & -\frac{27}{8}\left(N+4\right)^{2}\alpha\lambda^{3}+\frac{27}{32}\left(N+4\right)^{3}\lambda^{4}.
\end{align}
Then, we can find $D^{\left(5\right)}$ with the helps of $\beta$\textasciiacute s
and $\gamma'$s presents in\,(\ref{eq:D order equation}), as a results
we get

\begin{align}
D^{\left(5\right)} & =d_{1}\lambda\alpha^{3}\xi+d_{2}\alpha^{4}\xi+d_{3}\lambda^{3}\alpha\xi+d_{4}\alpha^{4}\lambda+d_{5}\alpha^{5}+d_{6}\lambda^{3}\alpha^{2}+d_{7}\lambda^{2}\alpha^{2}\xi+d_{8}\lambda^{2}\alpha^{3}\nonumber \\
 & +d_{9}\alpha\lambda^{4}+d_{10}\lambda^{5}+d_{11}\alpha^{6}\lambda^{-1}+d_{12}\alpha^{7}\lambda^{-2}+d_{13}\alpha^{8}\lambda^{-3}.
\end{align}
The coefficients $d_{1}$ to $d_{13}$ are presented in the appendix\,\ref{sec:All-coefficients-for D}.

These results will be used, in the next section, to study the modification
introduced by the leading logs summation in the DSB in our model.

\section{\label{sec:Dynamical-symmetric-breaking}Dynamical symmetric breaking}

In this section we will study the DSB in our model, for this, we will
use the results obtained in the previous section for the effective
potential up to five loops which was calculated using the renormalization
group equation, in the following form,

\begin{align}
V_{\mathrm{eff},R}^{\left(5\ell\right)}\left(\sigma\right) & =\sigma^{4}\left[A^{\left(1\right)}+\sum_{n=2}^{5}B^{\left(n\right)}L+\sum_{n=3}^{5}C^{\left(n\right)}L^{2}+\sum_{n=4}^{5}D^{\left(n\right)}L^{3}+\rho\right],\label{eq:Reg potential}
\end{align}
where $\rho$ is a finite renormalization constant and $V_{\mathrm{eff},R}^{\left(5\ell\right)}\left(\sigma\right)$
is the regularized effective potential up to five loops. The constant
$\rho$ is fixed using the CW normalization condition,
\begin{align}
\left.\frac{d^{4}}{d\sigma^{4}}V_{\mathrm{eff},R}\left(\sigma\right)\right|_{\sigma=\mu} & =\frac{4!}{4}\lambda.\label{eq:CW condition}
\end{align}

Requiring that $V_{\mathrm{eff},R}\left(\sigma\right)$ has a minimum
at $\sigma=\mu$ means imposing that 
\begin{align}
\left.\frac{d}{d\sigma}V_{\mathrm{eff},R}\left(\sigma\right)\right|_{\sigma=\mu} & =0,\label{eq:potential minimum}
\end{align}
which can be used to determine the value of $\lambda$ as a function
of free parameters $\alpha=e^{2},$ $\xi$ and $N$. Upon explicit
calculation, Eq.\,(\ref{eq:potential minimum}) turns out to be a
polynomial equation in $\lambda$, and among its solutions, we look
for real and positive values for $\lambda$, and correspond to a minimum
of the potential. i.e.,

\begin{align}
\left.\frac{d^{2}}{d\sigma^{2}}V_{\mathrm{eff},R}\left(\sigma\right)\right|_{\sigma=\mu} & >0,\label{eq:mass equation}
\end{align}
Using a program created in MATHEMATICA it was possible to verify for
which values of the free parameters $\alpha=e^{2}$, $\xi$ and $N$
we obtain a sensible value for $\lambda$, which means that the mechanism
of DSB is operational, and the symmetry is indeed broken by radiative
corrections.It is well-known that the effective potential can be gauge-dependent\,\citep{Jackiw1974}.
There is a sophisticated method to deal with this problem developed
by Nielsen\,\citep{Nielsen1975}, which for sake of simplicity, will
be properly addressed in a future work.

\begin{figure}
\begin{centering}
\includegraphics[width=0.8\columnwidth]{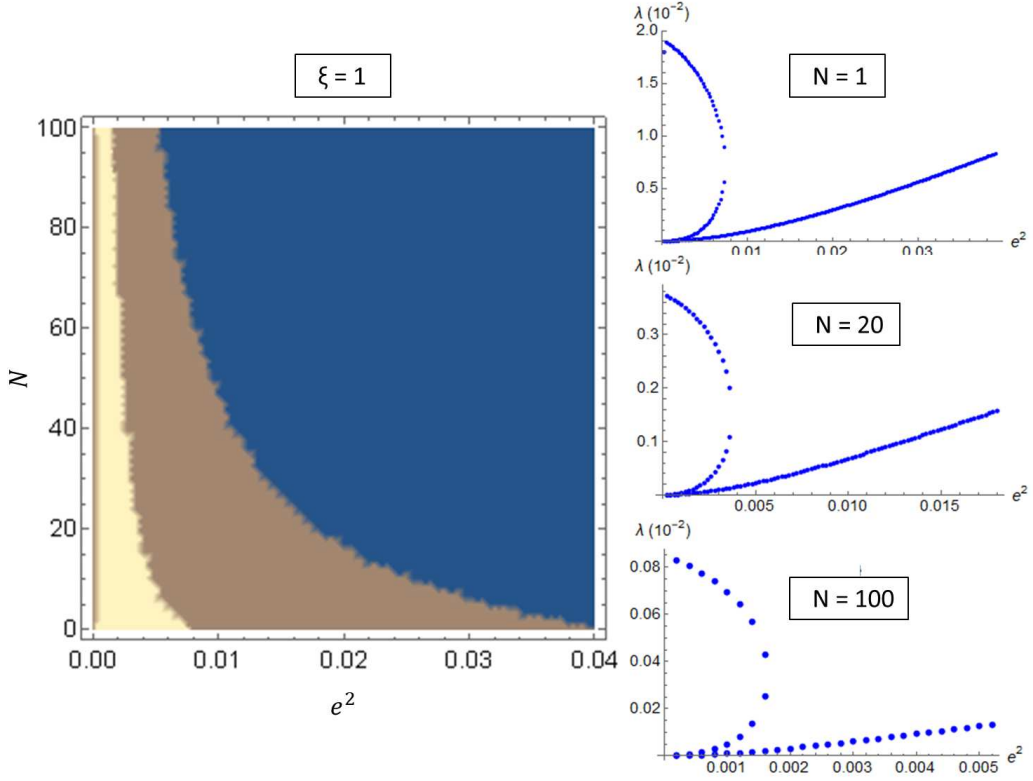}
\par\end{centering}
\centering{}\caption{\label{fig:Figura1}In the left hand side we show a region plot corresponding
to the result of scanning for DSB for different values of the free
parameters $e^{2}$ and $N$, with $\xi=1$. In our model we find
three regions: in the yellow region we have three possible solutions
for $\lambda$, in the brown one we have only one solution and in
the blue region we do not have solution for $\lambda$, meaning DSB
is not operational. In the right hand side, we show a set of plots
explicitly showing the behavior of the solutions for $\lambda$ as
a function of the $e^{2}$ parameter, for specific values of $N=1,20,100$
and $\xi=1$.}
\end{figure}

In order to suggest the rich structure of DSB in the model, we will
present some results for Feynman-t'Hooft gauge, i.e., $\xi=1$. We
considered $e^{2}$ and $N$ as free parameters, varying in the ranges
$0\leq e^{2}\leq0,04$ and $0\leq N\leq100$. Considering these values,
the parameter space in which the DSB occurs was analyzed, and the
summary of our findings is pictured in Fig.\,\ref{fig:Figura1}.
We notice the existence of three regions of solutions for $\lambda$,
where the yellow region is characterized by the existence of three
different solutions for $\lambda$, which means three different non-symmetric
vacua for each value of the parameters within this region. The brown
region corresponding to the existence of a single solution, and the
blue region with the absence of any solution which breaks symmetry,
i.e., for the case when DSB does not happen in our model.

\begin{figure}
\centering{}\includegraphics[scale=0.4]{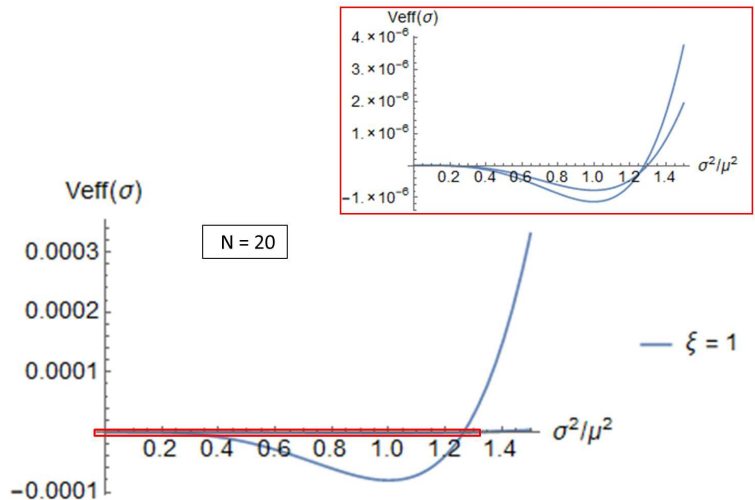}\caption{\label{fig:Figura2}This graph of $V_{\mathrm{eff}}\left(\sigma\right)$
vs. $\sigma^{2}/\mu^{2}$ shows the potential minimum for the gauge
parameter, $\xi=1$. The effective potential was evaluated using $e^{2}=0.001$,
$N=20$ and the values of $\lambda$, as can be seen in Eq.\,(\ref{eq:values of lambda}).
The red rectangle in the botton-left graph is shown in a different
scale in the top-right one.}
\end{figure}

We can analyze the minimum of the effective potential, Eq\,(\ref{eq:Reg potential}),
for example for values of $e^{2}=0.001$, $N=20$ and $\xi=1$, we
found the three values of $\lambda$,

\begin{eqnarray}
\lambda_{1}=0.003553,\,\, & \lambda_{2}=0.00003590,\,\, & \lambda_{3}=0.00001210,\label{eq:values of lambda}
\end{eqnarray}
and when these are used together we can see that the minimum occurs
as show in figure\,\ref{fig:Figura2}.

\section{\label{sec:Conclusion}Conclusion}

In this paper we have studied the behavior of effective potential
in a massless Abelian Higgs (AH) model with $N$-component complex
scalar field in $(3+1)$ dimensional space-time, specifically concerning
its classical vacua structure, obtaining hints of a very rich structure
of DSB, depending on the free parameters of the model (the gauge coupling
constant and the number of scalar fields). These results were obtained
by calculating the effective potential using the RGE equation, and
the $\widetilde{\beta}$ and $\widetilde{\gamma}$ functions already
reported in the literature. After adapting these renormalization group
functions, which were calculated in the minimal subtraction scheme,
to a renormalization scheme adequate for our purposes, we shown how
the RGE can be used to calculate, order by order in the logarithm
$L=\ln\left(\sigma^{2}/\mu^{2}\right)$ and the coupling constants,
the effective potential.

Our results points to some interesting prospects, that we intent to
approach in future publications. First, to investigate whether further
higher-order terms can be incorporated in the effective potential
by using a different summation, which could be implemented as a symbolic
package for MATHEMATICA, thus further improving our calculation (as
it was done in\,\citet{Quinto2016}). Second, a full study of the
Nielsen identity in our model would be important to factor out the
possible gauge dependence of the results. Finally, we expect to apply
this formalism for other models with application in condensed matter
and particle physics in other to study their DBS properties.
\begin{acknowledgments}
This work was supported by \emph{Fondo Nacional de Financiamiento
para la Ciencia, la Tecnología y la Innovación \textquotedbl Francisco
José de Caldas\textquotedbl , Minciencias }Grand No. 848-2019\emph{
}(AGQ), and Conselho Nacional de Desenvolvimento Científico e Tecnológico
(CNPq) Grant No. 305967/2020-7 (AFF).
\end{acknowledgments}

\appendix

\section{\label{sec:All-coefficients of B}All coefficients of $B$\textasciiacute s}

In this appendix we show the values of the coefficients associated
with the functions $B^{\left(3\right)}-B^{\left(5\right)}$ as a function
of the Riemann Zeta functions, $\zeta$. In our case we have $\zeta_{3}=\zeta\left(3\right)$
and $\zeta_{5}=\zeta\left(5\right)$ presents in our results. So,
we can fix these with $\zeta_{3}=1.202$, known as Apéry's constant,
and $\zeta_{5}=1.036$.
\begin{align}
b_{1}= & \frac{3}{2},\,\,\,b_{2}=\frac{25N}{8}+\frac{27}{2}+\frac{153}{2},\,\,\,b_{3}=\frac{3}{16}\left(47N+119\right),\,\,\,b_{4}=\frac{1}{16}\left(N\left(7N+10\right)-108\right),\nonumber \\
b_{5}= & \frac{3}{16}\left(N\left(5N+29\right)+57\right),\,\,\,b_{6}=-\frac{27}{4},\,\,\,b_{7}=\frac{63}{4},\nonumber \\
b_{8}= & \frac{1}{192}\left(452N^{2}+648\zeta_{3}\left(3N+5\right)+14433N+33633\right),\,\,\,b_{9}=\frac{21}{8}\left(N+4\right),\nonumber \\
b_{10}= & \frac{75\text{\ensuremath{\zeta_{3}}}}{4}+\frac{36599N^{2}}{1728}+\frac{7}{64}\left(673-48\text{\ensuremath{\zeta_{3}}}\right)N-\frac{3029}{32},\nonumber \\
b_{11}= & \frac{9}{4}-\frac{7N}{2},\,\,\,b_{12}=\frac{1}{384}\left(-112N^{3}+10541N^{2}-864\zeta_{3}\left(N+19\right)+56043N+139806\right),\nonumber \\
b_{13}= & \frac{1}{32}\left(84N^{3}+193N^{2}+72\zeta_{3}\left(N-1\right)-1721N-6432\right),\nonumber \\
b_{14}= & \frac{1}{128}\left(456N^{3}+3343N^{2}+144\text{\ensuremath{\zeta_{3}}}\left(5N+11\right)+8739N+11460\right),\nonumber \\
b_{15}= & -\frac{9\left(18N+109\right)}{8},\,\,\,b_{16}=\frac{459}{8},\,\,\,b_{17}=\frac{3}{16}\left(N\left(46N+257\right)+493\right),\nonumber \\
b_{18}= & \frac{3}{1920}\left(-204000\zeta_{3}+225600\text{\ensuremath{\zeta_{5}}}+608\pi^{4}+5404455\right)\nonumber \\
 & +\frac{1}{1920}N\left(-35400\zeta_{3}+43200\text{\ensuremath{\zeta_{5}}}+692\pi^{4}+7646330\right)\nonumber \\
 & +\frac{1}{1920}N^{2}\left(-30\left(1244\zeta_{3}+960\text{\ensuremath{\zeta_{5}}}-67285\right)+5N\left(70N+52113\right)+28\pi^{4}\right),\nonumber \\
b_{19}= & +\frac{5}{311040}N^{2}\left(34776N^{2}+1296\zeta_{3}(7N-1444)+11297005N+648\pi^{4}+47536876\right)\nonumber \\
 & \frac{4787\zeta_{3}}{8}-160\text{\ensuremath{\zeta_{5}}}-\frac{81}{311040}N\left(-222480\zeta_{3}+56\pi^{4}+1962495\right)\nonumber \\
 & -5\text{\ensuremath{\zeta_{5}}}N\left(N+21\right)+\frac{3\pi^{4}}{40}-\frac{2832569}{384},\,\,\,b_{20}=-\frac{3}{16}\left(5N\left(7N+32\right)+212\right),\,\,\,b_{21}=\frac{9}{2},\nonumber \\
b_{22}= & \frac{1}{34560}\left(5N^{2}\left(213840\zeta_{3}-51840\text{\ensuremath{\zeta_{5}}}+17(7987-144\text{\ensuremath{\zeta_{3}}})N-432\pi^{4}+2199556\right)\right)\nonumber \\
 & -\frac{216N}{34560}\left(-21330\zeta_{3}+18300\text{\ensuremath{\zeta_{5}}}+68\pi^{4}-173865\right)-\frac{135}{34560}\left(64944\zeta_{3}+183840\text{\text{\ensuremath{\zeta_{5}}}}+480\pi^{4}-1275293\right),\nonumber \\
b_{23}= & \frac{1}{48}\left(N\left(112N+633\right)+8163\right),\nonumber \\
b_{24}= & \frac{1}{17280}\left(4N^{2}\left(90\zeta_{3}(4N+1035)+57175N+162\pi^{4}+2051830\right)\right)\nonumber \\
 & +\frac{1}{17280}\left(81\left(43920\zeta_{3}+31200\text{\ensuremath{\zeta_{5}}}+132\pi^{4}+520405\right)+9N\left(5\left(78072\zeta_{3}-44640\text{\ensuremath{\zeta_{5}}}+895837\right)+228\pi^{4}\right)\right),\nonumber \\
b_{25}= & \frac{9}{8}\left(138-29N\right),\nonumber \\
b_{26}= & \frac{1}{3840}N\left(27360\text{\ensuremath{\zeta_{3}}}-72000\text{\ensuremath{\zeta_{5}}}+3N\left(8\left(2760\text{\ensuremath{\zeta_{3}}}+\pi^{4}-73135\right)+5N(4144N+10197)\right)-1008\pi^{4}-10087955\right)\nonumber \\
 & -\frac{24}{3840}\left(39720\text{\ensuremath{\zeta_{3}}}+9000\text{\ensuremath{\zeta_{5}}}+103\pi^{4}+779815\right),\nonumber \\
b_{27}= & \frac{1}{1280}\left(5N^{2}\left(9936\zeta_{3}-1920\text{\ensuremath{\zeta_{5}}}+4848N^{2}+46227N+8\pi^{4}+153516\right)+352\pi^{4}\right)\nonumber \\
 & +\frac{5}{1280}\left(86064\zeta_{3}-44640\text{\ensuremath{\zeta_{5}}}+226705\right)+\frac{2}{1280}N\left(155040\zeta_{3}-66000\text{\ensuremath{\zeta_{5}}}+124\pi^{4}+576805\right),\nonumber \\
b_{28}= & -\frac{333}{4},\,\,\,b_{29}=\frac{1}{96}\left(-N\left(8716N+139815\right)-372249\right),\,\,\,b_{30}=\frac{3}{16}\left(2207N+14742\right),\,\,\,b_{31}=-\frac{9639}{8}.
\end{align}

\section{\label{sec:All-coefficients-for C}All coefficients for $C$\textasciiacute s}

In this appendix we show the values of the coefficients associated
with the functions $C^{\left(4\right)}-C^{\left(5\right)}$,
\begin{align}
c_{1} & =\frac{27}{2},\,\,\,c_{2}=\frac{3}{32}\left(N\left(39N+967\right)+3069\right),\,\,\,c_{3}=\frac{1}{32}\left(N\left(298N+1551\right)+522\right),\nonumber \\
c_{4} & =\frac{1}{64}\left(N\left(N\left(7N+1255\right)+6897\right)+12042\right),\,\,\,c_{5}=\frac{9}{2}-\frac{21N}{8},\nonumber \\
c_{6} & =\frac{3}{16}\left(N\left(N\left(7N+50\right)+94\right)-102\right),\,\,\,c_{7}=\frac{9}{4}\left(N+4\right),\,\,\,c_{8}=\frac{9}{64}\left(N+4\right)\left(N\left(13N+51\right)+95\right),\nonumber \\
c_{9} & =-\frac{81}{8}\left(N+12\right),\,\,\,c_{10}=\frac{243}{8},\nonumber \\
c_{11} & =\frac{1}{2304}\left(N\left(9072\zeta_{3}+N\left(-32400\zeta_{3}+2\left(99446-21N\right)N+1170811\right)+980397\right)\right)\nonumber \\
 & +\frac{54}{2304}\left(8832\zeta_{3}-93709\right),\,\,c_{12}=\frac{9}{32}\left(\left(12-7N\right)N+112\right),\nonumber \\
c_{13} & =\frac{1}{384}\left(1366N^{3}+648\zeta_{3}\left(6N^{2}+45N+121\right)+89083N^{2}+631260N+1142397\right),\nonumber \\
c_{14} & =\frac{1}{128}\left(21N^{4}+11266N^{3}+80917N^{2}-72\zeta_{3}\left(N\left(17N+151\right)+516\right)+256995N+473436\right),\nonumber \\
c_{15} & =\frac{9}{16}\left(N\left(13N+82\right)+162\right),\nonumber \\
c_{16} & =\frac{3}{256}\left(686N^{4}+3268N^{3}-7937N^{2}+144\zeta_{3}\left(5\left(N-4\right)N-97\right)-74865N-151716\right),\nonumber \\
c_{17} & =270-\frac{69N}{16},\,\,\,c_{18}=\frac{9}{2},\nonumber \\
c_{19} & =\frac{1}{2304}\left(24743N^{3}+709337N^{2}+2592\zeta_{3}\left(5N\left(2N+15\right)-184\right)+2559492N+5545638\right),\nonumber \\
c_{20} & =\frac{1}{16}\left(N\left(35N+219\right)+3330\right),\nonumber \\
c_{21} & =\frac{9}{256}\left(270N^{4}+2573N^{3}+8763N^{2}+144\zeta_{3}\left(N+4\right)\left(5N+11\right)+14188N+14238\right),\nonumber \\
c_{22} & =-\frac{243}{4},\,\,\,c_{23}=-\frac{3}{32}\left(N\left(457N+9270\right)+35838\right),\,\,\,c_{24}=\frac{81}{16}\left(41N+340\right),\,\,\,c_{25}=-\frac{1215}{2}.
\end{align}

\section{\label{sec:All-coefficients-for D}All coefficients for $D$\textasciiacute s}

In this appendix we show the values of the coefficients associated
with the function $D^{\left(5\right)}$,
\begin{align}
d_{1} & =\frac{3}{8}\left(N\left(7N+36\right)+234\right),\,\,\,d_{2}=-\frac{135}{8}\left(N-6\right),\,\,\,d_{3}=\frac{27}{8}\left(N+4\right)^{2},\nonumber \\
d_{4} & =\frac{1}{192}\left(N\left(2N\left(616N+15255\right)+118449\right)+142074\right),\nonumber \\
d_{5} & =\frac{3}{32}\left(N\left(N\left(33N+1388\right)+11079\right)+23436\right),\nonumber \\
d_{6} & =\frac{3}{64}\left(N\left(N\left(7N\left(N+110\right)+5407\right)+12699\right)+14670\right),\nonumber \\
d_{7} & =-\frac{9}{16}\left(N+4\right)\left(7N+6\right),\,\,\,d_{8}=\frac{1}{192}\left(N\left(N\left(N\left(7N+5017\right)+38100\right)+106902\right)+58968\right),\nonumber \\
d_{9} & =\frac{9}{64}\left(N+4\right)\left(N\left(N\left(21N+142\right)+596\right)+376\right),\nonumber \\
d_{10} & =\frac{9}{64}\left(N+4\right)^{2}\left(N\left(23N+49\right)+77\right),\,\,\,d_{11}=-\frac{81}{8}\left(N\left(N+27\right)+129\right),\nonumber \\
d_{12} & =\frac{243}{4}\left(N+11\right),\,\,\,d_{13}=-\frac{729}{4}.
\end{align}

\bibliographystyle{unsrt}
\bibliography{Bibilografia}

\end{document}